# Response to Comment (arXiv: 1407.6854) on Palmer *et al.*, *Nature*, *510*, 385, 2014

by

Jeremy C. Palmer[§], Pablo G. Debenedetti[§], Roberto Car[¶] and Athanassios Z. Panagiotopoulos[§]

[§] Department of Chemical and Biological Engineering, Princeton University, Princeton, NJ 08544
[¶] Department of Chemistry, Princeton University, Princeton, NJ 08544

**Abstract**
We respond to a Comment [1] on our recent *Nature* paper [2]. We categorically disprove the arguments provided in [1] and thereby further substantiate the evidence presented in [2], demonstrating the existence of a metastable liquid-liquid transition in a molecular model of water. We will make our code publicly available shortly, along with proper user documentation that is currently under development.

**1. Introduction**
Recently we showed, using six state-of-the-art sampling techniques, that a molecular model of water exhibits a metastable liquid-liquid transition in the deeply supercooled regime [2]. In particular, we demonstrated the existence of three phases at the same, deeply supercooled condition, two of them liquids in metastable coexistence, as well as a stable crystal phase [2]. We further demonstrated full consistency between six free energy calculations, all of which showed a liquid-liquid transition at the same condition [2]. A recent Comment by Chandler [1] ignores this overwhelming evidence and attempts to discredit some of our calculations. This response examines systematically the claims made in [1] and categorically disproves them, thereby unambiguously reaffirming the evidence presented in [2]. Below we address each of the claims and arguments in [1], in the order in which they are presented.

**2. Binder's argument**
Chandler's Comment [1] begins by invoking an interesting argument by Kurt Binder [3]. Binder addresses the question of whether *critical fluctuations* can be detected in a metastable system due to nucleation of the stable phase. Binder argues against the possibility of observing critical singularities (*e.g.*, accurately measuring critical exponents) in a metastable state, because the finite lifetime of the metastable state would preclude the very large time necessary to equilibrate critical fluctuations. Binder's argument says absolutely nothing about the possibility of the existence of a metastable liquid-liquid transition, the topic of our *Nature* paper [2]. It says nothing about the free energy surface associated with such a metastable first-order transition (the main calculation in our paper, Fig. 1 in ref. [2]). It says nothing about the scaling of free energy barriers with system size, which metastability in no way alters. Our finite size scaling analysis of free energy barriers is not intended to investigate critical singularities: it has to do with demonstrating consistency with the existence of a first-order transition. We are able to perform such scaling analysis comfortably, without ever encountering crystallization problems. It is instructive to quote the last sentence of Binder's comment [3] (our italics) "This problem must prevent the observation of actual critical singularities, *even if liquid-liquid phase separation would occur.*" Thus, Chandler's attempt to invoke Binder's argument is incorrect.

It is also appropriate to point out that macroscopic samples of supercooled water have recently been *experimentally* shown to survive without crystallizing for millisecond time scales [4]. While such experiments do not address the hypothesized liquid-liquid transition (they were conducted at atmospheric pressure), they demonstrate that macroscopic samples [droplets of $O(10\mu m)$ in diameter] can be fully equilibrated down to 229 K. Thus, the *experimentally attained* ratio of the size of the fully relaxed (i.e., liquid) sample to the characteristic microscopic length is 50,000, far in excess of Chandler's estimate of $O(10)$. The experimentally observed large separation between crystallization and relaxation times under deeply supercooled conditions strongly suggests that the investigation of the hypothesized liquid-liquid transition should be feasible not just in simulations of small systems but in experiments performed on macroscopic samples.

**3. Free energy barrier scaling, Figure 1 of Chandler's Comment**
Figure 1 of Chandler's Comment [1] reports the system-size dependence of the free energy barrier separating the high-density and low-density liquids (HDL, LDL), using two sets of data: those from our *Nature* paper [2], and those from our *Faraday Discussion* paper [5] (ref. 7 in Chandler's Comment). Chandler writes that Palmer in a *Faraday Discussion* remark [6] (ref. 8 in Chandler's Comment) comments that the "results of Ref. 7 may require correcting but nothing is said about the origin of the correction in the new paper". This is a deliberately misleading statement because Palmer actually wrote (ref. [6], p. 136): "*We acknowledge in the original and current version of our Faraday Discussions manuscript that the finite size scaling calculation presented here does not provide definitive proof of a first-order liquid-liquid phase transition in ST2 water. Although we did not establish that each calculation was performed precisely at a point of liquid-liquid coexistence, finite-size effects on the coexistence pressure are expected to be relatively small. We have verified this by performing umbrella sampling calculations to demonstrate very clearly that the correct scaling behavior is obtained at conditions where coexistence can be established. I presented these results at the Faraday Discussion, and the calculations will be detailed in a forthcoming publication*". The above statement clearly indicates that the preliminary *Faraday* data were obtained *without establishing that each calculation was performed exactly at coexistence*. This condition was instead established for the *Nature* calculations [2].

It is thus profoundly misleading to say that our older results "*may require correcting but nothing is said about the origin of the correction in the new paper*" because the reason of the correction was well detailed in [6], as well as in [5] (page 92). It is also deeply misleading to report in the same figure preliminary data (the *Faraday* data), that were clearly presented as preliminary because precise coexistence had not been strictly established, and the definitive data (the *Nature* data), where coexistence is actually verified. The *Nature* data completely override the *Faraday* data, and do so according to clearly-documented procedures [2]. Reporting both is only done for the purpose of creating the impression that the uncertainty in our calculation is significantly larger than reported in the *Nature* paper. It is disappointing that Chandler persists in this argument about supposed inconsistencies between the *Faraday* and *Nature* free energy barrier calculations after we explicitly pointed out to him the above facts in e-mail communications on June 26 and July 23, 2014. These communications are available upon request.

The only purpose of using the *Faraday* data in Fig. 1 of Chandler's Comment [1], in other words, is to induce the impression of large uncertainties in our *Nature* paper [2] by invoking a supposed unexplained inconsistency that was in fact clearly explained [5, 6] long before the *Nature* paper was published.

Chandler's Comment [1] attempts to discredit our calculation by showing in his Fig. 1 that he can fit the data with a linear function of $N$. Everyone would agree that a finite size scaling analysis is not enough to *establish* without any doubt that the law has to be $N^{2/3}$. The basis for this scaling is theoretical, not numerical. It is hence sufficient for us to show that our data are consistent with the theoretical $N^{2/3}$ law. This we have clearly done, in addition to providing an overwhelming amount of fully consistent evidence that Chandler's Comment completely ignores.

Figure 1 below shows the results of an analogous exercise that we have carried out with the Limmer-Chandler crystallization free energy barrier data [7] for the mW coarse-grained water model [8]. It can be clearly seen that the data can be equally well fit with $N^{2/3}$ (the theoretical expectation), $N^{1/2}$, $N^{1/3}$ and $N^{3/4}$.

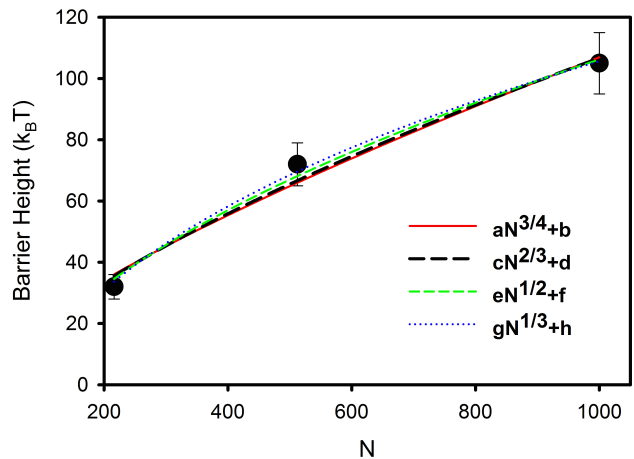

**Fig. 1: Fits to the Limmer-Chandler crystallization free energy barrier data (black circles) [7] for the mW coarse-grained water model [8]. Uncertainties were estimated from the symbol size in the inset of Fig. 5 in ref [7].**

Chandler's Comment [1] attempts to transform what one has every right to demand, namely that free energy barriers be *consistent with* theoretical expectations, as we have clearly done in our *Nature* paper (see Fig. 2 of ref. [2]), into the unattainable expectation that one *prove* that *only* the theoretical expectation, and no other, fit the data. Such a calculation is not only meaningless, but beyond current computational capabilities. Our calculations for the accessible system sizes yield free energy barriers that are consistent with the theoretically-expected $N^{2/3}$ scaling. Using this scaling, we have been able to extract HDL-LDL interfacial tensions that are fully consistent with a fluid-fluid transition, but inconsistent with a liquid-solid transition (see Methods in ref [2]).

The overwhelming weight of our evidence (ignored in Chandler's Comment) is the demonstration of the existence of three phases at the same, deeply supercooled temperature and pressure, two of which are in metastable equilibrium (Fig. 1 of our *Nature* paper [2]), and the fully consistent corroboration of a liquid-liquid transition using six state-of-the-art free energy techniques (Extended Data Figure 1 of the Methods section in our *Nature* paper [2]). It strains credulity that one can ignore this evidence and focus instead on the unattainable goal of proving numerically that *only* the theoretical expectation fits the data.

Contrary to Chandler's arguments, consistency of the free energy barrier with the theoretical exponent, rather than the stronger unique matching condition, does not change the fact that we have provided overwhelming evidence demonstrating a metastable liquid-liquid transition in a molecular model of water.

**4. Free energy profile, Figure 2 of Chandler's Comment**
Fig. 2 of Chandler's Comment [1] shows our free energy profile for $N$=600 ST2 water molecules. Chandler argues [1] that: "*The bottom of the left [LDL] basin is asymmetric, which is contrary to behavior expected of liquid matter. In particular, reversible fluctuations should yield a parabolic basin up to at least $k_BT$ above the minimum.*" This statement presumably arises from the fact that the LDL basin appears narrow in comparison to the HDL basin, but it is incorrect and trivial to falsify. Figure 2 below shows that a symmetric, parabolic function provides a good fit to the LDL basin up to ~2 $k_BT$ above the minimum. The maximum residual error of the parabolic fit within this region is less than 0.3 $k_BT$, which is significantly less than the uncertainties in the free energy, which are ~ $k_BT$. Thus, the parabolic nature of the LDL basin can be confirmed with relatively little effort. We also note that a version of Fig. 2 below was provided to Chandler on July 25, 2014, well in advance of the posting of his Comment [1]. Yet, his inaccurate claim remains intact.

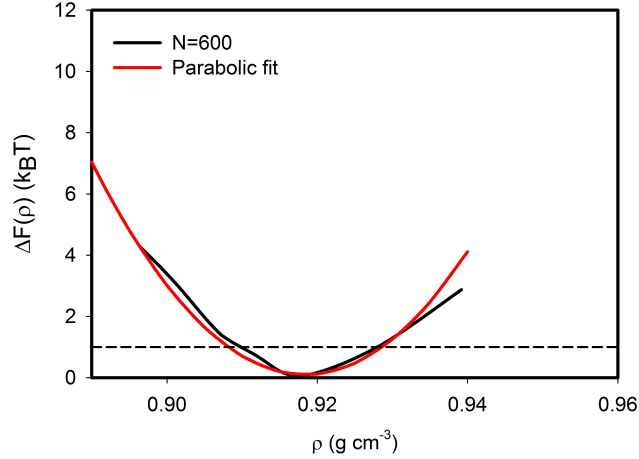

**Fig. 2: Parabolic fit to the LDL region of our free energy profile for *N*=600. The dashed reference line is positioned at *$k_B T$*.**

Chandler also claims [1]: "*The barrier separating the two basins [in Fig. 2 of ref 1] is relatively sharp, which is contrary to behavior expected in the presence of a stable interface… The only curvature at the barrier top should be due to the effects of finite size on fluctuations of the interface, and the barrier top should become flatter as the system size grows.*" This statement is technically correct, but it is grossly misleading because it fails to mention the fact that such flattening behavior is only expected to be prominent for relatively large systems away from the critical point, where interfacial fluctuations are small compared to the system's dimensions [9]. As expected, we observed relatively large interfacial fluctuations, which is consistent with the fact that our simulations were performed in relative proximity to the previously-estimated liquid-liquid critical temperature, at $T/T_c \approx 0.96$ [2], in order to avoid sampling problems associated with the sluggish relaxations at lower temperatures. Thus, as we have shown, our calculations are sufficient to satisfy consistency with the $N^{2/3}$ scaling law expected for a first-order liquid-liquid phase transition, but prominent flattening at the top of the barrier region is not observed. We instead find that the width of the barrier region increases with system size, which is a weaker manifestation of the flattening behavior described by Chandler [1]. This observation is entirely consistent with analogous free energy calculations that have been performed for the 3D Lennard-Jones fluid in similar proximity to the vapor-liquid critical point using systems of comparable size [9]. We also note that flattening behavior can be obscured when density is used as an order parameter [9], as was done in our *Nature* study [2].

**5. Relaxation times and autocorrelation functions**
In his Comment [1], Chandler argues that the time scale separation between density and bond-orientational order ($Q_6$) relaxation considered in his previous work [10] refers to the HDL region, and that our critique of Chandler's analysis in our *Nature* paper [2] is incorrect because we wrongly used relaxation data in the LDL region. He further argues that our autocorrelation functions show evidence of poor equilibration. Below we show categorically that both these claims are baseless, and we also address additional incorrect claims.

**a. Separation of time scales in the HDL and LDL regions**

Chandler's argument that we have wrongly used data from the LDL, instead of HDL region to disprove his hypothesis of time scale separation [10] is baseless: in fact, Limmer and Chandler invoked data from our previous calculations [11] in the LDL region to support their theory of "artificial polyamorphism". Specifically, Limmer and Chandler [10] report autocorrelation times for density and the bond-orientational order parameter $Q_6$ in Table I of ref [10] that were purportedly extracted from the autocorrelation functions in our 2012 study [11], which only presents such data for simulations performed in the LDL region. This data was juxtaposed with their calculations performed in the HDL region (*q.v.* Table I of ref [10]), and it was used to justify the main underlying assumption of their 2013 study [10], which claims that density relaxation occurs significantly faster than bond-orientational relaxation [10]. They further posit that such behavior leads to an illusory LDL-like basin that is not a metastable liquid, but rather a non-equilibrium artifact associated with the coarsening of ice [10]. The calculations reported in our *Nature* paper [2] show unambiguously that density and bond-orientational order are invariably coupled at long times, regardless of the technique used to perform the simulations [2]. Thus, we have falsified their separation of time scales hypothesis. Moreover, because they only observe a single amorphous basin in their free energy calculations, their definition of the HDL basin encompasses the low-density amorphous region we have examined.

**b. Imminent crystallization**

In our *Nature* study [2], we calculated a reversible free energy surface showing two liquid basins using long, unconstrained MC simulations that were run for ~100 $\tau_{Q_6}^{\text{LDL}}$ and showed no signs of crystallization, where $\tau_{Q_6}^{\text{LDL}}$ is the autocorrelation time for $Q_6$ in the LDL region. Chandler incorrectly argues [1] that these simulations "*contradict findings from 1 µs molecular dynamics trajectories that exhibit ice coarsening*", citing a recent paper by Yagasaki *et al.* [12]. In fact, these authors report finding evidence of a liquid-liquid transition in the ST2 model, which they claim is a distinct phenomenon from crystallization [12].

Chandler also states [1] that: "*The structural relaxation time for a high-density ST2 liquid at the conditions examined is about $10^2$ ps, and the relaxation time for $Q_6$ is about $10^4$ ps.*", and continues to argue that one-hundred times longer would reach $10^6$ ps, which is approximately the time where Yagasaki et al. [12] observed crystallization. Thus, he is claiming that it is not possible to run simulations for longer than ~100 $\tau_{Q_6}^{\text{HDL}}$ without observing crystallization. Not only have we demonstrated that this is possible [2], but we also find that it can routinely be achieved [2,6]. Chandler's argument fails to account for the fact that the characteristic crystallization time will decrease with system size. Yagasaki *et al.* [12] used 4,000 ST2 molecules, whereas our unconstrained calculations employ 192 water molecules. Since the simulations of ref [12] were run at thermodynamic conditions comparable to those of our *Nature* paper [2], the ratio of characteristic crystallization times should simply be inversely proportional to system size. The results of Yagasaki et al. [12] therefore suggest that crystallization should occur after

~$2 \times 10^7$ ps for the relatively small system we examined, which is more than an order of magnitude longer than claimed by Chandler [1].

**c. Autocorrelation functions, Figures 3 and 4 of Chandler's Comment**
Figure 4 of Chandler's Comment [1] shows autocorrelation functions for density and the bond-orientational order parameter $Q_6$ from our *Nature* study [2] that were computed from long, unconstrained MC simulations in the vicinity of the LDL region. Chandler makes two claims regarding these functions. He first argues that: "*At the shortest times for which they provide data, C(t) for the density shows a hint of two-step relaxation… The suggested early-time relaxation could reflect the relaxation processes that dominate at the higher densities, and it would be consistent with the time-scale separation between ρ and $Q_6$ shown to occur in the higher-density regime.*" This statement is purely speculative and pertains to the short time behavior, which is irrelevant to sampling because in our *Nature* study [2] we have computed our free energy surfaces by re-sampling our data over intervals significantly longer than the short time scales alluded to by Chandler.

Chandler's second claim is that the tails of our autocorrelation functions oscillate about a non-zero value, which he argues is indicative of systematic drift related to irreversible behavior [1]. Figure 4 of his Comment [1], however, clearly shows that our autocorrelation functions oscillate about zero shortly after exhibiting behavior where they become slightly anti-correlated. This is the typical behavior expected for a converged autocorrelation function. The long-time oscillatory behavior is statistically indistinguishable from zero because the uncertainty in the tails of the correlation functions is relatively large, ranging from 0.1 to 0.2 in this region. This is a consequence of the fact that our particular autocorrelation functions shown in Chandler's Comment [1] were computed by averaging over $O(10)$ unconstrained simulations that remained within the vicinity of the LDL basin for their entire duration, as stated in our *Nature* paper [2]. Consequently, relatively few sub-trajectories are included in the tail region of these particular autocorrelation functions. Thus, the tails are statistically indistinguishable from zero and Chandler's argument is incorrect. We note that several of the autocorrelation functions reported in our *Nature* paper [2] were run for times that are orders of magnitude longer than the unconstrained simulations referred to by Chandler (see Extended Data Table 1 of ref [2].)

The irony of Chandler's second claim is that the Boltzmann-weighted average $Q_6$ autocorrelation function shown for his simulation data in Fig. 3 of his Comment [1] clearly exhibits oscillations in the tail region about a non-zero value. This function is computed using $\langle C(t) \rangle = \sum_i w_i C_i(t)$, where $w_i$ is the Boltzmann weight for the i$^{th}$ umbrella sampling window in the HDL region. Because $w_i \geq 0$, the weights preserve the sign of the individual correlation function. It is therefore impossible for the average correlation function to oscillate around a negative value unless the individual correlation functions also exhibit this behavior. We expect that this would be noticeable in Fig. 3 of Chandler's Comment [1] if the individual $Q_6$ autocorrelation functions had not been obscured by plotting far many lines on the same graph.

## 6. Freezing

In our *Nature* paper [2] we noted order-of-magnitude discrepancies between our calculated free energy difference between HDL and ice Ic and the value reported in Fig. 5b of Chandler's previous publication [10]. The Methods section of our *Nature* paper accordingly includes detailed calculations aimed at clarifying the origin of this discrepancy and its important thermodynamic consequences [2]. These calculations [2] clearly demonstrate that our data withstand thermodynamic scrutiny [2], whereas Chandler's [10] don't. These calculations naturally involve freezing behavior, given that what is in question here is a free energy difference between a liquid and a crystal. In his Comment [1], Chandler essentially argues melting behavior is not an important aspect of understanding the phase behavior of supercooled water, and attempts to minimize the discrepancy between our respective free energy differences by providing a very different numerical value for this quantity than the one we read directly from his published work. Below we show that the number provided by Chandler [1] is meaningless, since it pertains to a version of the ST2 model that we did not study; and we demonstrate that the freezing data provided by Chandler [1] contradict his published claims [10] as to the accuracy of his free energy surfaces, and contains such large error bars as to render these calculations irrelevant to the analysis of phase behavior.

Chandler argues that we "*report [in ref [2]] that coexistence conditions for the ST2 model are known, but that is not true.*" In fact, we have confirmed that the estimate of ice Ih's melting temperature by Weber and Stillinger [13] referenced in our *Nature* paper [2] is accurate. We have also computed melting curves for ice Ic for two variants of the ST2 model (ST2b and ST2c in ref [10]) using three different techniques. Results from these calculations, which will be presented in a forthcoming publication, show unambiguously that Limmer and Chandler's [10] estimates of ice Ic's melting temperature are highly inaccurate for these models, predicting melting temperatures that are too low by ~40 K at the reported pressures. Although in his Comment Chandler claims not to have calculated ice Ic-liquid coexistence, he is apparently unaware that his own Fig. 13 in ref [10] (top row, middle column for ST2b at 2.6 kbar) shows a liquid and a crystal phase with the same free energy at the given temperature and pressure, that is to say, equilibrium melting conditions. Chandler's main argument [1] against our interpretation of his data [10] for the ST2 model we examined in our *Nature* study [2] (*i.e.,* ST2b in ref [10]) is that his data for this model "*is insufficient to locate a phase coexistence point*". As explained above, phase equilibrium has in fact been calculated by Chandler [10].

Comparison of Fig. 1 in our *Nature* paper [2] and Fig. 5b in ref [10] reveals a very large, order-of-magnitude discrepancy between the ice Ic-liquid free energy differences at comparable conditions. We find that ice Ic is lower in free energy by -742 J/mol (229 K and 2.4 kbar), whereas Limmer and Chandler report -66 J/mol (230 K and 2.2 kbar). In our *Nature* paper, we explore in detail the implication of this large discrepancy, and we document that our calculations withstand thermodynamic scrutiny, whereas Chandler's do not (see Extended Data Tables 2 and 3 in ref [2]). Chandler attempts to dismiss this criticism by stating that his "*best estimate*" of this free energy difference is -400 J/mol, which would indeed remove a large part of our discrepancy. There is only one problem with this argument: Chandler's "*best estimate*" pertains to a different variant of the ST2 model (ST2a) than the one we used (ST2b) and is therefore irrelevant. Furthermore, in his Comment [1], Chandler does not explain how his "*best estimate*" for the ST2a model was obtained, and he also does not provide the relevant state conditions for which this value is supposedly relevant. From Figs. 2, 5 and 13 in ref [10] we obtain, for the ST2a model, ice Ic-liquid free energy differences of approximately -53 J/mol at 235 K and 2.2 kbar, and -160, -106, and -66 J/mol at 230 K and 1.8, 2.2 and 2.6 kbar, respectively.

Interestingly, Chandler [1] does not explain why the ST2a model is his "*best estimate*". We argue that this is a result of the fact that the ST2a model was the only variant actually simulated in Chandler's previous studies [7,10]. Free energy surfaces for the ST2b and ST2c models (Figs 5b and 5c in ref [10], respectively) were derived from the data for the ST2a model by performing single-stage free energy perturbation calculations via Eq. (8) in ref [10]. This fact has been confirmed by the lead author of ref [7,10]. Monte Carlo simulations were therefore only directly performed for ST2a, a variant for which a liquid-liquid phase transition has not been found by any research group, as indicated in Table 2 of ref [10]. Thus, the free energy surfaces for the two ST2 variants (ST2b and ST2c), for which a liquid-liquid transition has been located [2,14], were obtained indirectly from the ST2a model using a free energy perturbation method that is rigorous in principle, but notoriously prone to error due to the fact it can exhibit poor convergence, even in cases where the free energy difference between the reference and final state is relatively small [15].

We therefore posit that Limmer and Chandler's free energy perturbation approach resulted in large statistical uncertainties (and possibly significant systematic errors) in the results shown for the ST2b and ST2c variants in their 2013 study [10]. This is evidenced by the ~1 kbar uncertainty in the coexistence pressure for the ST2c model reported in Fig. 5b of ref [1], which demonstrates that the free energy calculations are not converged to within sufficient accuracy to provide relevant information regarding the existence of a liquid-liquid transition. To put this 1 kbar uncertainty in perspective, the pressures where HDL and LDL reach their respective limits of stability are only separated by ~0.6 kbar at 235 K for ST2c [14]. For ST2b at 228.6 K, this pressure difference is reduced to ~0.4 kbar [11]. In other words, the uncertainty in the calculation exceeds the magnitude of the pressure range where the relevant physics occurs. We reiterate that, contrary to Chandler's assertions, melting calculations are essential to address serious matters of thermodynamic consistency. This is what we have done.

**7. Conclusion**
This Response disproves the claims made in Chandler's Comment [1]. It thereby further strengthens the conclusions of our recent *Nature* paper [2], which firmly establishes the existence of a metastable liquid-liquid transition in a molecular model of water. Although the criticisms leveled in [1] are easily dismissed, the origin of the very significant differences between our calculations [2] and those of Limmer and Chandler [7,10] remain unexplained to this date. We have shown in [2] that the results presented in [10] *for the* *same version of the ST2 model that we have used* do not withstand thermodynamic scrutiny.

We also suggest that in order to clarify this discrepancy, Chandler must perform actual Monte Carlo simulations for the exact variants of ST2 (*i.e.*, ST2b and ST2c in ref [10]) for which others have reported observing a liquid-liquid phase transition [2,14]. As it stands, the only free energy surfaces that Chandler has reported for the aforementioned variants of ST2 have been derived not by direct simulation, but by applying free energy perturbation to the data for ST2a [10], for which such transition has not been reported. It is therefore not entirely surprising that Chandler's results to our knowledge are the only published free energy calculations contradicting the existence of a liquid-liquid transition in relevant variants of ST2.

We will make one of the numerous advanced sampling codes that we have developed for our *Nature* study [2] publicly available at http://pablonet.princeton.edu/pgd/html/links.html soon after we complete the writing of user documentation. This will happen no later than September 1, 2014. We believe that transparency is necessary for resolving this long-standing controversy, and we sincerely hope that Chandler will follow our lead and make his hybrid Monte Carlo code available to the scientific community.

**Acknowledgments**
P.G.D. acknowledges support from the National Science Foundation (CHE 1213343). R.C. and A.Z.P. acknowledge support from the US Department of Energy (DE-SC0008626 and DE-SC0002128, respectively).

**References**
[1] D. Chandler, arXiv:1407.6854 (2014).
[2] J. C. Palmer, F. Martelli, Y. Liu, R. Car, A. Z. Panagiotopoulos and P. G. Debenedetti, *Nature*, *510*, 385 (2014).
[3] K. Binder, *PNAS*, *111*, 9374 (2014).
[4] J. A. Sellberg *et al.*, *Nature*, 510, 381 (2014).
[5] J. C. Palmer, R. Car and P. G. Debenedetti, *Faraday Discuss.*, *167*, 77 (2013).
[6] J. C. Palmer, General Discussion, *Faraday Discuss.*, *167*, 118 (2013).
[7] D. T. Limmer and D. Chandler, *J. Chem. Phys.*, *135*, 134503 (2011).
[8] V. Molinero and E.B. Moore, *J. Phys. Chem. B*, *113*, 4008 (2009).
[9] J. E. Hunter and W. P. Reinhardt , *J. Chem. Phys.*, *103*, 8627 (1995).
[10] D.T. Limmer and D. Chandler, *J. Chem. Phys.*, *138*, 214504 (2013).
[11] Y. Liu, J. C. Palmer, A. Z. Panagiotopoulos and P. G. Debenedetti*, J. Chem. Phys.*, 137, 214505 (2012).
[12] T. Yagasaki, M. Matsumoto and H. Tanaka, *Phys. Rev. E*, 89, 020301 (2014).
[13] T. A. Weber and F. H. Stillinger, *J. Chem. Phys.,* 80, 438 (1984).
[14] P. H. Poole*,* R. K. Bowles, I. Saika-Voivod and F. Sciortino, *J. Chem. Phys.*, 138, 034505 (2013)
[15] C. Chipot and A. Pohorille, *Free energy calculations : theory and applications in chemistry and biology* (Springer, Berlin ; New York, 2007).